\documentstyle[12pt]{article}
\newcommand {\beq}{\begin{equation}}
\newcommand {\eeq}{\end{equation}}
\newcommand {\beqa}{\begin{eqnarray}}
\newcommand {\eeqa}{\end{eqnarray}}
\newcommand {\n}{\nonumber \\}

\begin{document}
\setlength{\oddsidemargin}{0cm}
\setlength{\baselineskip}{7mm}

\begin{titlepage}
 \renewcommand{\thefootnote}{\fnsymbol{footnote}}
$\mbox{ }$
\begin{flushright}
\begin{tabular}{l}
KEK-TH-736 \\
Dec. 2000
\end{tabular}
\end{flushright}

~~\\
~~\\
~~\\

\vspace*{0cm}
    \begin{Large}
       \vspace{2cm}
       \begin{center}
         {High Energy Behavior of Wilson Lines}
\\
       \end{center}
    \end{Large}

  \vspace{1cm}

\begin{center}

           Avinash D{\sc har}
\footnote{ On leave from Dept of Theoretical Phys, Tata Institute,
Mumbai 400005, INDIA.}
\footnote
           {
e-mail address : adhar@post.kek.jp},
           Yoshihisa K{\sc itazawa}\footnote
           {
e-mail address : kitazawa@post.kek.jp}

         {\it Laboratory for Particle and Nuclear Physics,\\
High Energy Accelerator Research Organization (KEK),}\\
               {\it Tsukuba, Ibaraki 305-0801, Japan} \\
\end{center}

\vfill

\begin{abstract}
\noindent
We investigate the high energy behavior of the
correlation functions of the open Wilson lines
in noncommutative gauge theory.
We obtain a very simple physical picture that
they are bound to form a group of closed Wilson loops.
We prove our claim in the weak coupling region
by perturbative analysis. We emphasize
the importance of respecting the cyclic
symmetry of the straight Wilson lines
to compute the correlation functions.
The implications for stringy calculation
of the correlators are also discussed.
\end{abstract}

\vfill
\end{titlepage}
\vfil\eject

\section{Introduction}
\setcounter{equation}{0}
\setcounter{footnote}{0}

In matrix model formulation of superstring theory,
not only matter but also space\cite{BFSS}
or even space-time\cite{IKKT} may emerge out of
matrices.
Noncommutative gauge theory may be regarded as a concrete realization
of such a possibility\cite{CDS}.
We can obtain fully noncommutative gauge theory by expanding
IIB matrix model around noncommutative backgrounds\cite{AIIKKT}.
\footnote{
NCYM is equivalent to large $N$ twisted reduced model
\cite{RM}\cite{twisted}\cite{nakayama}. See also \cite{Li}.}
In fact such a matrix model construction
has been very useful to elucidate physical properties
of noncommutative gauge theory\cite{GMS}.
It has further been argued that
four dimensional noncommutative gauge theory
with maximal SUSY ($NCYM_4$) may be interpreted as superstring
theory with noncommutativity scale as its effective
string scale\cite{IIKK2}.
It may serve as a testing ground for the proposal that matrix models may
serve as nonperturbative formulation of superstring theory.

The gauge invariant observables in noncommutative
gauge theory involve not only closed Wilson loops
but also open Wilson loops (Wilson lines)
\cite{IIKK}\cite{AMNS}\cite{Gross}\cite{Dhar}\cite{Rey}.
It has also been pointed out that the operators
which couple to graviton multiplets may be
constructed through them\cite{IIKK}.
The validity of such an argument is confirmed
by the recent investigations
\cite{Liu}\cite{Trivedi}.
These authors have also shown that
the Wilson lines are closely related to
the Seiberg-Witten map\cite{SW}.

The correlators of the Wilson lines
are found to exhibit stringy exponential
suppression behavior at high energy limit
\cite{Gross}.
In our recent work, we have investigated
the strong coupling behavior of the
expectation value of the Wilson loops\cite{DK}.
Our proposal is to consider Nambu-Goto action in the metric of the
corresponding supergravity solution.
The four dimensional part of the metric is conformally flat while
the conformal factor possesses the unique maximum in the
radial coordinate.
We postulate that the Wilson loops in $NCYM_4$ may be
represented by
the closed contours at the the maximum of the space-time metric.
In our proposal, the expectation value of the Wilson loops
are given by the extremum of the Nambu-Goto action,
namely the minimum area.
We have further mentioned that the high energy limit
of the Wilson line correlators may be identical
to the expectation value of the large Wilson loops.

In this paper we construct a proof of the above stated
equivalence.
We first give an heuristic argument based on the loop
equations in section 2. In section 3,
we prove the equivalence by the
perturbative analysis of the correlators in the weak coupling region.
Since our argument based on the loop equation holds
in both the weak and strong coupling regions, we expect that
the equivalence does not break down even in the strong coupling regime.
If so, our recent work in \cite{DK} becomes relevant to this problem.
In the concluding section,
we discuss possible strong coupling behavior of
the Wilson line correlators in the high energy limit
based on such a line of argument.

\section{Loop equations in NCYM}
\setcounter{equation}{0}

In this section,
we briefly review matrix model
constructions of NCYM first\cite{AIIKKT}.
We recall IIB matrix model action:
\beq
S  =  -{1\over g^2}Tr({1\over 4}[A_{\mu},A_{\nu}][A^{\mu},A^{\nu}]
+{1\over 2}\bar{\psi}\Gamma ^{\mu}[A_{\mu},\psi ]) .
\label{action}
\eeq
Here $\psi$ is a ten dimensional Majorana-Weyl spinor field, and
$A_{\mu}$ and $\psi$ are $N \times N$ Hermitian matrices.
The noncommutative Yang-Mills theory is obtained by expanding the
theory around the following background:
\beq
[\hat{p}_{\mu},\hat{p}_{\nu}]=iB_{\mu\nu} ,
\eeq
where  $B_{\mu\nu}$ are $c$-numbers.
We assume the rank of $B_{\mu\nu}$
to be $\tilde{d}$ and define its inverse $C^{\mu\nu}$ in $\tilde{d}$
dimensional subspace.
We expand $A_{\mu}=\hat{p}_{\mu}+\hat{a}_{\mu}$.
Noncommutative Yang-Mills can be realized through matrix models by the
the following map from matrices onto functions
\beqa
\hat{a} &\rightarrow& a(x) ,\n
\hat{a}\hat{b}&\rightarrow& a(x)\star b(x) ,\n
Tr&\rightarrow&
\sqrt{det B}({1\over 2\pi})^{\tilde{d}\over 2}\int d^{\tilde{d}}x,\n
g^2&\rightarrow&\sqrt{det B}({1\over 2\pi})^{\tilde{d}\over 2}g_{NC}^2 .
\label{momrule}
\eeqa

The gauge invariant observables in NCYM are the Wilson loops:
\beq
W(C)=Tr[Pexp(i\int_C d x^{\mu}(\sigma )A_{\mu})]
\eeq
where $C$ denotes a contour parametrized by $x^{\mu}(\sigma )$.
Let us consider the following correlation functions:
\beq
\int dAd\psi Tr[t^aPexp(i\int_{C_1} d x^{\mu}(\sigma )A_{\mu})]
W(C_2)e^{-S} .
\eeq
We consider the following infinitesimal change of the variables as
$A_{\mu}\rightarrow A_{\mu}+\epsilon t^a$.
By using the completeness condition of the generators of $U(N)$:
\beq
\sum_a t^a_{ij}t^a_{kl}=\delta_{il}\delta_{jk} ,
\eeq
we obtain,
\beqa
&&<{1\over g^2}Tr[\{[A_{\mu},[A_{\mu},A_{\nu}]]
+{1\over 2}\Gamma_{\nu}[\psi,\bar{\psi}]_{+}\}
Pexp(i\int_{C_1} d x^{\mu}(\sigma_1 )A_{\mu})]W(C_2)>\n
&=&i\int_0^1 dt \dot{x}^{\nu} <Tr[Pexp(i\int_{0}^t d\sigma_1
\dot{x}^{\mu}A_{\mu})] Tr[Pexp(i\int_{t}^1 d\sigma_1
\dot{x}^{\mu}A_{\mu})]W(C_2)>\n &+&i\int_0^1 ds \dot{x}^{\nu}
<Tr[Pexp(i\int_{0}^s d\sigma_2 \dot{x}^{\mu}A_{\mu})
exp(i\int_{C_1} d x^{\mu}(\sigma_1 )A_{\mu})
exp(i\int_{s}^1 d\sigma_2 \dot{x}^{\mu}A_{\mu})]> .\n
\label{loopeq}
\eeqa
The first and second term on the right-hand side of the above equation
represents the splitting and joining of the loops respectively
\cite{FKKT}.

The Wilson loops in the matrix model are mapped to
those in NCYM by eq.(\ref{momrule}):
\beqa
&&Tr[Pexp(i\int_C d x^{\mu}(\sigma )A_{\mu})]\n
&=&
\sqrt{det B}({1\over 2\pi})^{\tilde{d}\over 2}
\int d^{\tilde{d}}xPexp(i\int_C d
y^{\mu}(\sigma )a_{\mu}(x+y(\sigma ))) exp(ik^{\mu}x_{\mu}))_{\star}
\label{ncloop}
\eeqa
where the total momentum $k_{\mu}$ is related to
the vector $d^{\mu}$ which connects the two ends of the contour
as $k_{\mu}=B_{\mu\nu}d^{\nu}$.
The symbol $\star$ in the above expression reminds us that
all products of fields must be understood as $\star$ products.
In order to prove the above identity
we consider the following discretization of a Wilson loop.
\beq
Pexp(i\int_C d x^{\mu}(\sigma )A_{\mu})
=\lim_{n\rightarrow \infty}\prod_{j=1}^n exp(i\Delta x_j\cdot
(\hat{p}+\hat{a})) .
\eeq
The above quantity can be rewritten as
\beqa
&&\lim_{n\rightarrow \infty}\prod_{j=1}^n exp(i\Delta x_j\cdot\hat{p})
exp(i\Delta x_j\hat{a})\n
&=&\lim_{n\rightarrow \infty}(\prod_{j=1}^n
V_jexp(i\Delta x_j\hat{a})V_j^{\dagger})V_n\n
&=&\lim_{n\rightarrow \infty}(\prod_{j=1}^n
exp(i\Delta x_jV_j\hat{a}V_j^{\dagger}))V_n
\eeqa
where $V_j=(\prod_{k=1}^j
exp(i\Delta x_k\hat{p})$ is the translation operator along the contour
$C$ from the first to the $j$-th segment.
After applying eq.(\ref{momrule}) to the above
and taking $n\rightarrow \infty$
limit of it, we obtain the right-hand side of eq.(\ref{ncloop}).
We then obtain the Schwinger-Dyson equations in NCYM\cite{Dorn}
\beqa
&&<\int d^{\tilde{d}}x[{1\over i}\{[D_{\mu},[D_{\mu},D_{\nu}]]
+{1\over 2}\Gamma_{\nu}[\psi,\bar{\psi}]_{+}\}\n
&&\times Pexp(i\int_{C_1} d y^{\mu}(\sigma_1 )a_{\mu}(x+y(\sigma))
exp(ik_1\cdot x))]_{\star}W(C_2)>\n
&=&g_{NC}^2{det B}({1\over 2\pi})^{\tilde{d}}
\int_0^1 dt \dot{x}^{\nu}
<\int d^{\tilde{d}}x_1[Pexp(i\int_{C} d y^{\mu}(\sigma_1
)a_{\mu}(x_1+y(\sigma_1)) exp(ik\cdot x_1))]_{\star}\n
&&\times\int d^{\tilde{d}}x_2
[Pexp(i\int_{C_1-C} d y^{\mu}(\sigma_2 )a_{\mu}(x_2+y(\sigma_2))
exp(i(k_1-k)\cdot x_2))]_{\star}W(C_2)>\n
&+&g_{NC}^2\int_0^1 ds \dot{x}^{\nu}
<\int d^{\tilde{d}}x[Pexp(i\int_{\tilde{C}} d y^{\mu}(\sigma_1
)a_{\mu}(x+y(\sigma_1)) exp(i\tilde{k}\cdot x))\n
&&\times
Pexp(i\int_{C_1} d y^{\mu}(\sigma_1
)a_{\mu}(x+y(\sigma)))exp(ik_1^{\mu}x_{\mu}))
\n&&\times
Pexp(i\int_{C_2-\tilde{C}} d y^{\mu}(\sigma_2 )a_{\mu}(x+y(\sigma_2))
exp(i(k_2-\tilde{k})\cdot x))]_{\star}> .
\label{sdeqn}
\eeqa

We may generalize the above equation in $U(1)$ gauge group to $U(m)$
by considering the direct product of noncommutative space-time and
gauge group which can be realized in $U(Nm)$.
We consider the generators of $U(Nm)$ which satisfy
\beq
\sum_a \sum_bt^a_{ij}\lambda^b_{\alpha\beta}t^a_{kl}\lambda^b_{\gamma\delta}
=\delta_{il}\delta_{jk}\delta_{\alpha\delta}\delta_{\beta\gamma}
\eeq
where $\lambda^b$ denotes the generators of $U(m)$.
The Wilson loop becomes
\beqa
&&{1\over m}Tr[Pexp(i\int_C d x^{\mu}(\sigma )A_{\mu})]\n
&=&
{1\over m}\sqrt{det B}({1\over 2\pi})^{\tilde{d}\over 2}
\int d^{\tilde{d}}x tr[Pexp(i\int_C d
y^{\mu}(\sigma )a_{\mu}(x+y(\sigma ))) exp(ik^{\mu}x_{\mu})]_{\star} .
\label{mcloop}
\eeqa

Following the analogous steps as in $U(1)$ case, we obtain
\beqa
&&<{1\over m}\int d^{\tilde{d}}xtr[{1\over i}\{[D_{\mu},[D_{\mu},D_{\nu}]]
+{1\over 2}\Gamma_{\nu}[\psi,\bar{\psi}]_{+}\}\n
&&\times Pexp(i\int_{C_1} d y^{\mu}(\sigma_1 )a_{\mu}(x+y(\sigma))
exp(ik_1\cdot x))]_{\star}W(C_2)>\n
&=&\lambda{det B}({1\over 2\pi})^{\tilde{d}}
{1\over m^2}\int_0^1 dt \dot{x}^{\nu}
<\int d^{\tilde{d}}x_1tr[Pexp(i\int_{C} d y^{\mu}(\sigma_1
)a_{\mu}(x_1+y(\sigma_1)) exp(ik\cdot x_1))]_{\star}\n
&&\times\int d^{\tilde{d}}x_2
tr[Pexp(i\int_{C_1-C} d y^{\mu}(\sigma_2 )a_{\mu}(x_2+y(\sigma_2))
exp(i(k_1-k)\cdot x_2))]_{\star}W(C_2)>\n
&+&{\lambda\over m^2}\int_0^1 ds \dot{x}^{\nu}
{1\over m}<\int d^{\tilde{d}}xtr[Pexp(i\int_{\tilde{C}} d y^{\mu}(\sigma_1
)a_{\mu}(x+y(\sigma_1)) exp(i\tilde{k}\cdot x))\n
&&\times
Pexp(i\int_{C_1} d y^{\mu}(\sigma_1
)a_{\mu}(x+y(\sigma)))exp(ik_1^{\mu}x_{\mu}))
\n&&\times
Pexp(i\int_{C_2-\tilde{C}} d y^{\mu}(\sigma_2 )a_{\mu}(x+y(\sigma_2))
exp(i(k_2-\tilde{k})\cdot x))]_{\star}>
\label{sdeqnum}
\eeqa
where $tr$ denotes the trace operation over $U(m)$
and $\lambda=g_{NC}^2m$ is the 't Hooft coupling.

The correlation functions of the Wilson lines
can be investigated in principle by solving these
loop equations.
In this paper we show that the correlators of the
very long Wilson lines
are identical to the expectation value of the Wilson loop
which can be formed from them.
Since very long Wilson lines carry large momenta, it is the
equivalence which holds in the high energy limit.
In fact such an equivalence can be expected from the loop equations.

Let us consider the simplest looking loop equation eq.(\ref{loopeq}).
The first term and the second term on the right hand side
represent splitting and joining terms respectively.
For long straight Wilson lines, we can argue that the joining
term is dominant since it is proportional to the length
of $C_2$.
Since the left-hand side of the equation is
essentially the deformation of the
first Wilson line $C_1$, we can argue that the
expectation value of Wilson lines is equivalent to that of completely joined
Wilson lines namely closed Wilson loop.
In the next section,
we construct a perturbative proof of the equivalence
which is in accord with the loop equation argument.

\section{Wilson lines as Wilson loops}
\setcounter{equation}{0}

In NCYM, we also have
the gauge invariant operators specified by the open contours
as is defined in eq.(\ref{ncloop}).
Let us consider the two point function of $W(C_1)$ and $W(C_2)$
for example. Two contours $C_1$ and $C_2$ specify the
shapes of two open segments. Let us denote the vector which
connects the two ends of a contour by $\vec{d}$.
In noncommutative Yang-Mills, such an open segment carries
the momentum $\vec{k}_{\mu}=B_{\mu\nu}\vec{d}^{\nu}$.
In order for two point functions to be nonvanishing,
$W(C_1)$ and $W(C_2)$ must possess vanishing total momentum.
In other words, $C_1$ and $C_2$ can be put together to form
a closed loop. Let us denote the vector which connects
the tail of $C_1$ and head of $C_2$ as $\vec{x}$.
What is calculated
in noncommutative Yang-Mills theory can be expressed as
\beq
<W(C_1)W(C_2)>
=V\int d^4x exp(i\vec{k}\cdot \vec{x})
<W(C_1,C_2,\vec{x})> .
\eeq
Namely it can be interpreted in terms of
the expectation value of the operator
specified by a closed loop which
contains two segments $C_1$ and $C_2$.
However $<W(C_1,C_2,\vec{x})>$
cannot be obtained by a simple Fourier transformation of
$<W(C_1)W(C_2)>$,
since the contours $C_1$ and $C_2$
depend on $\vec{k}$.

In fact we can express the two point function
of the Wilson loops in terms of the one point
function through matrix model constructions.
If the gauge group is $U(1)$, we obtain
\beqa
&&Tr[Pexp(i\int_{C_1} d x^{\mu}(\sigma )A_{\mu})]
Tr[Pexp(i\int_{C_2} d x^{\mu}(\sigma )A_{\mu})]\n
&=&{1\over N}\sum_{\vec{l}}
Tr[Pexp(i\int_{C_1} d x^{\mu}(\sigma )A_{\mu})
exp(i\vec{l}\cdot \hat{p})
Pexp(i\int_{C_2} d x^{\mu}(\sigma )A_{\mu})
exp(-i\vec{l}\cdot \hat{p})]
\label{u(1)2pt}
\eeqa
where we have used the completeness condition
of the generators of $U(N)$.
\beq
{1\over N}\sum_{\vec{l}}exp(i\vec{l}\cdot \hat{p})_{a,b}
exp(-i\vec{l}\cdot \hat{p})_{c,d}
=\delta_{a,d}\delta_{b,c} .
\eeq
If the gauge group is $U(m)$ , we either keep the
trace operation over $U(m)$ in each Wilson line
or insert the generators of $U(m) (\lambda^a )$ such as
\beqa
&&Tr[Pexp(i\int_{C_1} d x^{\mu}(\sigma )A_{\mu})]
Tr[Pexp(i\int_{C_2} d x^{\mu}(\sigma )A_{\mu})]\n
&=&\sum_a{1\over N}\sum_{\vec{l}}
Tr[Pexp(i\int_{C_1} d x^{\mu}(\sigma )A_{\mu})
\lambda^aexp(i\vec{l}\cdot \hat{p})\n
&&\times Pexp(i\int_{C_2} d x^{\mu}(\sigma )A_{\mu})
\lambda^aexp(-i\vec{l}\cdot \hat{p})] .
\eeqa
The matrix model constructions such as eq.(\ref{u(1)2pt}) can be
translated into noncommutative gauge theory expressions as follows:
\beqa
&&exp(i\Phi)\int d^{4}{x}
\int d^{4}\tilde{x} exp(ik^{\mu}\tilde{x}_{\mu})\n
&&[Pexp(i\int_{C_1} d y_1^{\mu}(\sigma )a_{\mu}(x+y_1(\sigma )))
Pexp(i\int_{C_2} d y_2^{\mu}(\sigma )a_{\mu}(x+\tilde{x}+y_2(\sigma ))
)]_{\star}
\label{phase}
\eeqa
where $\Phi$ is the magnetic flux enclosed in the closed loop
$C_1+C_2$.
It is given by the area of the loop
divided by $C$.

We next consider the three point function.
\beqa
&&Tr[Pexp(i\int_{C_1} d x^{\mu}(\sigma )A_{\mu})]
Tr[Pexp(i\int_{C_2} d x^{\mu}(\sigma )A_{\mu})]
Tr[Pexp(i\int_{C_3} d x^{\mu}(\sigma )A_{\mu})]\n
&=&{1\over N}\sum_{\vec{l}}{1\over N}\sum_{\vec{m}}
Tr[Pexp(i\int_{C_1} d x^{\mu}(\sigma )A_{\mu})
exp(i\vec{l}\cdot \hat{p})
Pexp(i\int_{C_2} d x^{\mu}(\sigma )A_{\mu})\n
&&\times
exp(i\vec{m}\cdot \hat{p})
Pexp(i\int_{C_2} d x^{\mu}(\sigma )A_{\mu})
exp(-i\vec{m}\cdot \hat{p})
exp(-i\vec{l}\cdot \hat{p})] .
\label{u(1)3pt}
\eeqa
It can be translated into noncommutative gauge theory as
\beqa
&&exp(i\Phi)\int d^{4}x
\int d^{4}x_2\int d^{4}x_3exp(i(k_2+k_3)\cdot x_{2}+ik_3\cdot x_3)\n
&&[Pexp(i\int_{C_1} d y^{\mu}(\sigma )a_{\mu}(x+y(\sigma )))
exp(i\int_{C_2} d y^{\mu}(\sigma )a_{\mu}(x+x_2+y(\sigma )))\n
&&\times exp(i\int_{C_3} d y^{\mu}(\sigma )a_{\mu}(x+x_2+x_3+y(\sigma ))
)]_{\star}
\label{3point}
\eeqa
where $x_2$ connects the tail of $C_1$ and head of $C_2$ and
so does $x_3$ between $C_2$ and $C_3$.
In the case of $n$ point functions of the Wilson lines,
the prescription is just analogous. What is calculated
in noncommutative gauge theory is
\beqa
&&<W(C_1)W(C_2)\cdots W(C_n)>\n
&=&V\prod_{i=1}^n\int d^4x_i exp(i\sum_{i=2}^{i=n}\vec{l}_i\cdot \vec{x}_i)
<W(C_1,C_2,\cdots ,C_n;\vec{x}_1,\vec{x}_2,\cdots ,\vec{x}_{n-1})> .
\eeqa
In this expression
$x_1$ denotes the center of mass coordinate of the system.
The vector $\vec{x}_i (i\geq2) $ connects the tail of $C_{i-1}$ and the
head of $C_{i}$ and $l_i=\sum_{j=i}^{j=n}k_j$.

The real space correlator can be regarded as a one point
function of the operator of a single closed loop.
The loop consists of the segments $C_i$ which are accompanied with the gauge
fields. These segments are connected by straight segments of length
$|\vec{x}_i|$ which do not involve gauge fields.
Although these are new types of the operators
which are distinct from closed Wilson loops, they are not
gauge invariant observables by themselves.
Here we would like to point out that such an object is useful to
discuss the correlation functions
of long Wilson lines which carry large momenta.
In such a region, we can show that the correlators of the Wilson
lines are indistinguishable from those of large closed Wilson loops.
In what follows we prove our contention in the weak coupling region by using
the perturbation theory.

Let us consider the leading contribution
to the two point function of straight
Wilson lines.
\beqa
&&\int d^{4}x
\int d^{4}x_1exp(ik\cdot x_1)\n
&&<P \int_{C_1} d a\cdot A(x+x_1+a)
\star \int_{C_2} d {b}_1\cdot A(x+b)>\n
&=&V\int dadb exp(-ik\cdot (a-b)){g_{NC}^2\over k^2}
\label{2ptamp}
\eeqa
where $V$ denotes the space-time volume.
$a$ and $b$ denote the locations of the two ends
of a propagator on the Wilson lines $C_1$ and $C_2$
respectively.
We also use $A(x)$ to denote gauge
fields in the remaining of this section.
This expression is invariant under
independent translations of $a$
and $b$ along the contours of the Wilson lines.
It is the manifestation of the
cyclic symmetry of the straight Wilson lines.
Due to this symmetry we can fix $a=b=0$ after factoring out
the volume factor $Ck_1Ck_2$.
Since $x_1\sim 1/k$, we find that the lowest order contribution
represents a long rectangle of the horizontal length $Ck$
and the vertical length of $1/k$.

We move on to
the three point functions in eq.(\ref{3point}).
In the weak coupling region, we may retain
\beqa
&&exp(i\Phi)\int d^{4}x
\int d^{4}x_1\int d^{4}x_3exp(ik_1\cdot x_1+ik_3\cdot x_3)\n
&&<P \int_{C_1} d a\cdot A(x+x_1+a)
\star \int_{C_2} d {b}_1\cdot A(x+b_1)\n
&&\star \int_{C_2} d {b}_2\cdot A(x+{b}_2)
\star \int_{C_3} d c\cdot A(x+x_3+c )>
\label{3ptamp}
\eeqa
where we have only considered the amplitude where
gluons are exchanged between $(C_1,C_2)$ and
$(C_2,C_3)$. We also need to add three other types
of contributions to obtain the full amplitude at the
tree level.
In this particular example, we have
not only the ordered contributions with $b_1<b_2$ but also
the other ordering $b_2<b_1$ where by $b_1$ we imply
the location of the propagator with momentum $k_1$.
The latter contribution can be
reexpressed as
\beqa
&& exp(ik_3\cdot ({x}+b_2))
\star exp(ik_1\cdot ({x}+b_1)) \n
&=&exp(ik_1\cdot ({x}+b_1))\star
exp(ik_3\cdot ({x}+b_2+Ck_2 )) .
\eeqa
In this expression $\tilde{b}_2=b_2+Ck_2$ is such that
$b_1<\tilde{b}_2<b_1+Ck_2$. It can be
shown to be independent of $b_1$ after changing the variable
from $b_2$ to $\tilde{b}_2$.

We note here that eq.(\ref{3ptamp}) is invariant
under independent translations of $a$ along $C_1$ and $c$ along $C_3$.
It is also invariant under the simultaneous
translations of $b_1,b_2$ along $C_2$.
Due to the symmetry, we can fix the location of $a,b_1,c$ at the
ends of the Wilson line after factoring out the volume factor
$Ck_1Ck_2Ck_3$.
In this way, we obtain
\beqa
&&V g_{NC}^4exp({i\over
2}k_1Ck_2sin(\theta_1))cos(\theta_1)cos(\theta_3){1\over
{k_1}^2}{1\over {k_3}^2} Ck_1Ck_2Ck_3 \n &&
\int_0^{Ck_2} d|b_2|exp(-ik_1sin(\theta_1)|b_2|)\n
&=&V {g_{NC}^4}cot(\theta_1)cos(\theta_3)
{1\over {k_1}^2}{1\over {k_3}^2}
{1\over i}CCk_2 Ck_3\n
&&\{exp({i\over 2}k_1Ck_2sin(\theta_1))
-exp(-{i\over 2}k_1Ck_2sin(\theta_1))\}
\label{lowest}
\eeqa
where $\theta_1$ and $\theta_3$ are the angles of the two
corners of the triangle formed by $(C_1,C_2)$ and $(C_2,C_3)$
respectively.

We emphasize here that
straight Wilson lines possess the cyclic symmetry.
In fact the Wilson lines can be written as
$Tr(U^n)$ in the matrix model construction
where $U=exp(i\Delta x\cdot(\hat{p}+\hat{a}))$
and $\Delta x=Ck/n$.
In order to evaluate the correlation functions
which involve a particular Wilson line,
we need to attach propagators with momenta $\{p_i\}$ to it.
It is clear that there are $n$ ways to pick the first
gauge field with momentum $p_1$ from $Tr(U^n)$.
They are all equivalent due to the cyclic symmetry of the trace.
If we connect two propagators, only the relative position of the
second one to the first matters due to the cyclic
symmetry. So in general the correlator is proportional
to the length of the Wilson line and we can fix the location of
one of the propagators.

The propagators are very short
in the large $k$ limit since
their lengths are of $O(1/k)$.
Furthermore
the $|b_2|$ integration is only supported
by the two infinitesimal segments of
the width $O(1/k)$ at the
boundaries of the integration region.
Therefore in the large $k$ limit,
the first term on the right-hand side of eq.(\ref{lowest})
corresponds to a closed triangle
configuration made of $(C_1,C_2,C_3)$ in that order.
The second term appears to represent the configuration where three
Wilson lines share the same end point.
However we need to recall the cyclic symmetry of the Wilson lines.
The points on the same Wilson line can
be freely moved simultaneously. In the case of
two points on the same Wilson line, a pair of vanishingly close points
$(b_1,b_2)$
is equivalent to the pair of the end points $(b_2,b_1)$
due to the cyclic symmetry.
In this way we can interpret it as the closed triangle made of $(C_2,C_1,C_3)$
in that order.
Therefore we can indeed see that the Wilson lines are bound to form
closed Wilson loops in the large $k$ limit in the lowest order
of perturbation theory.
In the three point function case, we find two different triangles which can be
formed by the Wilson lines.
We can also understand the phase in each term
$\pm k_1Ck_2sin(\theta_1)/2$
as the magnetic flux passing through the respective triangle.
Although there is a topologically distinct diagram which
involves the three point vertices at the tree level,
the structure of such a contribution can be shown to be just analogous.

We can extend similar analysis to $n$ point functions of
the Wilson lines. By such an analysis we can easily convince ourselves that
the $n$ point functions of Wilson lines
are effectively described by a group of Wilson loops
which can be formed from the Wilson lines.
It is certainly clear that the correlation function is
saturated by a finite numbers of configurations for
$n$ point functions just like the three point function case.
We argue that
they could only be Wilson loops due to the gauge invariance.

In the weak coupling region, we might argue that the investigation
of the tree diagrams suffices.
However it is certainly not so for two point functions\cite{Gross}.
It is found that the leading contribution at the $n$-th order is
$(\lambda |k| |Ck|/4\pi )^n/(n!)^2$.
The summation over $n$ can be estimated by the saddle
point method as
$exp(\sqrt{{\lambda} |k| |Ck|/\pi})$.
The average separation of the two
Wilson lines is found to be $<n>/k\sim \sqrt{\lambda |Ck|/|k|}$.
Although it is much larger than the tree level estimate $1/k$,
it is still much smaller than the noncommutativity scale
in the weak coupling regime.

In the case of three point functions, we also find
logarithmic divergences in association of the
corners. The first example occurs at the next order.
\beqa
&&exp(i\Phi)\int d^{4}x
\int d^{4}x_1\int d^{4}x_3exp(ik_1\cdot x_1+ik_3\cdot x_3)\n
&&P \int_{C_1} d a\cdot A(x+x_1+a_2)\star \int_{C_1} d a\cdot A(x+x_1+a_1)\n
&&\star \int_{C_2} d {b}_1\cdot A(x+b_1)
\star \int_{C_2} d {b}_2\cdot A(x+{b}_2)
\star \int_{C_2} d {b}_3\cdot A(x+{b}_3)
\star \int_{C_3} d c\cdot A(x+x_3+c )\n
&=&exp(i\Phi)\int d^{4}x
\int d^{4}x_1\int d^{4}x_3exp(ik_1\cdot x_{1}+ik_3\cdot x_3)\n
&&\times \int_{C_1}d a_1\cdot \int_{C_2} d b_1
\int_{C_1}d a_2\cdot \int_{C_2} d b_2
\int_{C_2} d b_3\cdot \int_{C_3} d c\n
&&{g_{NC}^2\over 4\pi^2 (x_1+a_1-b_1)^2}
{g_{NC}^2\over 4\pi^2(x_1+a_2-b_2)^2}
{mg_{NC}^2\over 4\pi^2(x_3+c-b_3)^2}\n
&=&V {\lambda^3\over m^2} exp(i\Phi)\int_{C_1} d a_1\cdot \int_{C_2} d b_1
\int_{C_1} d a_2\cdot \int_{C_2} d b_2
\int_{C_2} d b_3\cdot \int_{C_3} d c\n
&&\int {d^4p\over (2\pi )^4}{1\over (k_1-p)^2}{1\over p^2}{1\over {k_3}^2}
exp(-i(k_1-p)\cdot (a_1-b_1)-ip\cdot (a_2-b_2)-ik_3\cdot (c-b_3)) .\n
\eeqa
We can now fix $a_1=b_1=c=0$ due to the cyclic symmetry of the Wilson lines.
\beqa
&&V {\lambda^3\over m^2}exp(i\Phi)cos(\theta_1)cos(\theta_3)Ck_1Ck_2Ck_3
\int_{C_1} d a_2\cdot \int_{C_2} d b_2
\int_{C_2} d |b_3|\n
&&\int {d^4p\over (2\pi )^4}{1\over (k_1-p)^2}{1\over {p}^2}{1\over {k_3}^2}
exp(-ip\cdot (a_2-b_2)+ik_3\cdot b_3)\n
&=&V i{\lambda^3\over m^2}exp(i\Phi)cos^2(\theta_1)cot(\theta_3)Ck_1Ck_2C
\int d|a_2|d|b_2|
\int {d^4p\over (2\pi )^4}{1\over (k_1-p)^2}{1\over {p}^2}{1\over {k_3}^2}\n
&&exp(-ip\cdot (a_2-b_2))
(1-exp(ik_3\cdot (Ck_2+b_2)))
\n
&=&
V {i\lambda^3\over m^2}exp(i\Phi)cos^2(\theta_1)cot(\theta_3)Ck_1Ck_2C
\int d|a_2|d|b_2|
\int {d^4p\over (2\pi )^4}{1\over (k_1-p)^2}{1\over {p}^2}{1\over {k_3}^2}\n
&&\{exp(-ip\cdot (a_2-b_2))\n
&&-exp(i(k_1-p)\cdot (a_2-b_2)-ik_1\cdot Ck_2)\}\n
&=&
V {\lambda^3\over m^2}exp(i\Phi)cos^2(\theta_1)cot(\theta_3)Ck_1Ck_2C
\int d|b_2|
\int {d^4p\over (2\pi )^4}{1\over
(k_1-p)^2}{1\over p^2}{1\over {k_3}^2}\n
&&\{exp(ip\cdot b_2)(exp(-ip\cdot Ck_1)-1){Ck_1\over p\cdot Ck_1 }\n
&&+exp(-i(k_1-p)\cdot b_2-ik_1\cdot
Ck_2)(exp(i(k_1-p)\cdot Ck_1)-1){Ck_1\over (k_1-p)\cdot Ck_1 }\}\n
&=&
V {\lambda^3\over im^2}cos^2(\theta_1)cot(\theta_3)Ck_1Ck_2C
\int {d^4p\over (2\pi )^4}{1\over
(k_1-p)^2}{1\over {p}^2}{1\over {k_3}^2}\n
&&\{exp({i\over 2}k_1\cdot Ck_2){Ck_1Ck_2\over p\cdot Ck_1p\cdot Ck_2}
(1-exp(-ip\cdot Ck_1))(1-exp(-ip\cdot Ck_2))\n
&&-exp(-{i\over 2}k_1\cdot Ck_2){Ck_1Ck_2\over (k_1-p)\cdot Ck_1(k_1-p)\cdot
Ck_2}\n &&\times(1-exp(i(k_1-p)\cdot
Ck_1))(1-exp(i(k_1-p)\cdot Ck_2))\} .
\eeqa

In the first term of the above expression,
we notice the following factor
\beq
\int {d^4p\over (2\pi )^4}{Ck_1Ck_2\over {p}^2p\cdot Ck_1 p\cdot Ck_2} .
\eeq
Due to the additional propagator $1/(k_1-p)^2$ in the full expression,
the large momentum cut-off scale is $k_1$.
The small momentum cut-off can also be seen to be $O(1/Ck)$.
Therefore it could give rise to a large factor
of $O(\lambda log(Ck^2))$ in the large $k$ limit.
It can be regarded as the correction to the first term
of the tree amplitude in eq.(\ref{lowest}).
It can be associated
with a corner of the first triangle.
The second term is just analogous after the change of the variables
from $(k_1-p)$ to $p$.
It is the correction to the second term in eq.(\ref{lowest}).
It can also be associated with a corner of the second triangle.
Since each loop gives rise to an additional logarithmic factor
in the ladder diagrams, we need to sum them to all orders
in the leading log approximation.

For this purpose
we consider a generic ladder diagram with $n$ propagators
around a corner of a triangle
\beqa
&&\int d^{4}x
\int d^{4}x_1exp(ik_1\cdot x_1)\n
&&P \int_{C_1} d a_n\cdot A(x+x_1+a_n)\star
\cdots \star \int_{C_1} d a_2\cdot A(x+x_1+a_2) \star \int_{C_1}d a_1\cdot
A(x+x_1+a_1)\n
&& \star \int_{C_2} d {b}_1\cdot A(x+b_1)\star \int_{C_2}d {b}_2\cdot A(x+b_2)
\star \cdots \star \int_{C_2}d {b}_n\cdot A(x+b_n)\n
&&\cdots\n
&=&
{1\over m}\int d^{4}x
\int d^{4}x_1exp(ik_1\cdot x_1)\n
&&\times \int_{C_1}d a_1\cdot \int_{C_2} d b_1
\int_{C_1}d a_2\cdot \int_{C_2} d b_2
\cdots \int_{C_1}d a_n\cdot \int_{C_2} d b_m\n
&&{\lambda\over 4\pi^2 (x_1+a_1-b_1)^2}{\lambda\over 4\pi^2 (x_1+a_2-b_2)^2}
\cdots {\lambda\over 4\pi^2 (x_1+a_n-b_n)^2}\n
&&\cdots\n
&=&V{\lambda^{n}\over m}\int_{C_1} d a_1\cdot \int_{C_2} d b_1
\int_{C_1} d a_2\cdot \int_{C_2} d b_2 \cdots
\int_{C_1} d a_n\cdot \int_{C_2} d b_n\n
&&\int{d^4p_1\over (2\pi)^4 }{exp(ip_1\cdot(a_1-b_1))\over {p_1}^2}
\int{d^4p_2\over (2\pi)^4 } {exp(ip_2\cdot(a_2-b_2))\over {p_2}^2}
\cdots \int{d^4p_n\over (2\pi)^4 }{exp(ip_n\cdot(a_n-b_n))\over {p_n}^2} \n
&&(2\pi)^4\delta (k_1-\sum_ip_i)\n
&&\cdots .
\label{ladder}
\eeqa

Here we may fix $a_1=b_1=0$ by the cyclic symmetry of the Wilson line.
In the leading log approximation,
we consider the contributions from the phase space where
the momenta are strongly ordered.
\beqa
&&k_1\sim p_1>>p_2>>\cdots >>p_n .
\label{stgodr}
\eeqa
In real space, it corresponds to the following region
\beqa
&& 1/k_1 \sim x_1 << a_2-b_2 << \cdots << a_n-b_n .
\label{rspodr}
\eeqa
Let us assume that we get logarithmic factors for
each leg of the ladder as we will find shortly.
Let us assume that the scale of the legs of the ladder
is uniformly distributed in the logarithmic scale over $log(L)$.
We can then estimate the amplitude with $n$ legs
as $(log(L)/n)^n \sim log^n(L)/n!$.
Therefore these characteristic log factors in the leading log approximation
imply that the scale of the legs is uniformly distributed
in the logarithmic scale.
The uniform distribution in the logarithmic scale corresponds to
strongly ordered distributions in phase space eq.(\ref{stgodr})
or in real space eq.(\ref{rspodr}).
With such an approximation, the above expression can be evaluated as
\beqa
&&{\lambda^{n}}cos(\theta_1)^n
{1\over k_1^2}Ck_1Ck_2\n
&&
\int_{|a|_1}^{Ck_1} d |a|_2\int d |b|_2
{1\over 4\pi^2 (a_2-b_2)^2}
\cdots \int_{|a|_{n-1}}^{Ck_1}d |a|_n\int d |b|_n
{1\over 4\pi^2 (a_n-b_n)^2}\n
&&\cdots .
\eeqa

Our strategy is to perform integrations over $b$ next.
Although we have to deal with nested integrations,
the problem simplifies due to the strong ordering.
Let us consider a particular leg of the ladder.
Then the inner legs are much shorter and the outer legs
are much longer than it.
So effectively we can shrink all the inner legs to the point
and move the outer legs to the infinity.
Therefore we are left with the following single integral
\beqa
\int d|b| {1\over (a-b)^2}
&=& \int _{-|a|cos(\theta_1)}^{\infty}db
{1\over |a|^2sin(\theta_1)^2+|b|^2}\n
&=&{1\over |a|sin(\theta_1)}
\int _{-cot(\theta_1)}^{\infty}db
{1\over 1+b^2}\n
&=& {1\over |a|sin(\theta_1)}(\pi-\theta_1) .
\eeqa

After integrating over $|b|$ variables
in this way, we obtain
\beqa
&&({\lambda\over 4\pi^2})^{n}cot(\theta_1)^{n-1}
cos(\theta_1)(\pi-\theta_1)^{n-1}
{1\over k_1^2}Ck_1Ck_2\n
&&
\int_{1/k_1}^{Ck_1} d |a_2|
{1\over |a_2|}
\cdots \int_{|a_{n-1}|}^{Ck_1}d |a_n|
{1\over |a_n|} \n
&&\cdots .
\eeqa
Since $|x|<|a_2|<\cdots <|a_n|<k_1$, the integration ranges
are easy to understand.
With these justifications, the above integral is found to be
\beqa
&&({\lambda\over 4\pi^2})cos(\theta_1)
{1\over k_1^2}Ck_1Ck_2\n
&&({\lambda\over 4\pi^2}cot(\theta_1)(\pi-\theta_1)
log(Ck_1^2))^{n-1}{1\over (n-1)!} \n
&&\cdots .
\label{logfac}
\eeqa

After summing over $n$ , we obtain the power
enhancement factor in association with a corner
\beqa
exp({\lambda cot(\theta_1)(\pi-\theta_1)\over 4\pi^2 }log({Ck_1^2})) .
\eeqa
It is clear that such a power law can be associated
with each corner of the Wilson loops formed by the
Wilson lines.
\footnote{
The possibility of such a power law enhancement
between nearly parallel Wilson lines is
independently noted by \cite{Rozali}.}
Such corner divergences are
well known to occur in the Wilson loop
expectation values.
It is also studied
in ordinary gauge theory through
$AdS$/CFT correspondence\cite{Ooguri}.
The appearance of such a power law enhancement
in the Wilson line correlators in the high energy limit
which is characteristic
to the Wilson loops is consistent with our assertion that
they are indeed equivalent.

\section{Conclusions and Discussions}
\setcounter{equation}{0}
In this paper we have shown that high energy behavior of
the correlators of the Wilson lines is
identical to the expectation value of large Wilson loops.
In the case of three point functions,
the dominant region of the phase space in eq.(\ref{stgodr}) corresponds to
the configuration such that $x_1=x_3=1/k$ since the propagators which
are localized at the two corners of the triangle are of the
length of $1/k$. The Wilson lines effectively form
a closed loop in such a region of the phase space.
Therefore the length of the segments which do not involve gauge fields
are vanishingly small. So we expect that their effect is equivalent
to insertion of local operators in the closed Wilson loops.
If so, they do not influence the universal
large momentum behavior of the correlators.
These arguments also apply to multi-point functions of Wilson lines.
We therefore find that the high energy limit of the Wilson
line correlators can be described by large closed
Wilson loops in the weak coupling region.

In the high energy limit in the weak coupling regime,
the multi-point correlation functions of the normalized Wilson lines
behave as
\beqa
&&<W(k_1)W(k_2)\cdots W(k_n)>\n
&\sim&exp(-\sum_{i=1}^{i=n}\sqrt{{\lambda\over 4\pi}|C k_i||k_i|}
+\sum_{i=1}^{i=n}{\lambda cot(\theta_i)(\pi-\theta_i)\over 4\pi^2}log(|C
k||k|))
\label{weak}
\eeqa
where $\theta_i$ is the $i$-th angle of the relevant closed
Wilson loop and we assume that $k_i$ are all of the same order $k$.
In this expression, the exponential suppression
of the normalized multi-point function is
caused by the exponential enhancement of the
two point function.

It is conceivable that the equivalence of the Wilson lines
and Wilson loops at high energy continues to hold at
strong coupling.
We may be able to prove it by making
the loop equation argument precise.
Here we would like to refer to our work where the
Wilson loop expectation value in NCYM has been investigated
by Nambu-Goto action\cite{DK}.
There we consider superstring theory in a particular background.
The string frame metric possess the maximum at the
scale $R\sim (\lambda)^{1/4}$ in the fifth radial coordinate $r$.
We have proposed to put the Wilson loops at $r=R$.
In such a construction, we obtain analogous
expression with $AdS$/CFT correspondence
since the relevant Wilson loops are large.
The only novelty is to
identify the short-distance cut-off with $R$.
With this prescription we predict the strong coupling behavior of the
Wilson lines as follows:
\beqa
&&<W(k_1)W(k_2)\cdots W(k_n)>\n
&\sim&exp(-R\sum_{i=1}^{i=n}\sqrt{|C k_i||k_i|}
+R^2\sum_{i=1}^{i=n}{cot(\theta_i){(\pi-\theta_i)}\over \pi}log(|C
k||k|/R^2)) .
\label{strong}
\eeqa
It is because $\sum_{i=1}^{i=n}\sqrt{|C k_i||k_i|}$
is proportional to the perimeter length of the
Wilson loop in fully noncommutative gauge theory.

We recall here that the average distance of the two point function
is $O(R^2)$ with respect to the noncommutativity scale
in the weak coupling limit as it is explained in section 3.
The standard prescription $R^2\rightarrow R$ may imply that
the minimal length scale which can be probed by the
high energy limit of the two point function is indeed $R$\
in the strong coupling limit.
It is hence likely that
our prescription to put the Wilson loops
at $r=R$ is relevant
in such a limit.
The $log(|Ck||k|)$ behavior in the weak coupling
expression in eq.(\ref{weak})
is already remarkable in that the short distance cut-off
$1/k$ and the long distance cut-off $Ck$ are related through
the noncommutativity scale $C$.
It is reminiscent of $T$ duality
in string theory.
However eq.(\ref{strong}) predicts
the appearance of $R^2$ factor in the logarithm
in the strong coupling limit.

\begin{center} \begin{large}
Acknowledgments
\end{large} \end{center}
This work is supported in part by the Grant-in-Aid for Scientific
Research from the Ministry of Education, Science and Culture of Japan.

\end{document}